\documentclass{article}

\usepackage[preprint, nonatbib]{nips_2018}

\usepackage{float}

\usepackage[square,numbers]{natbib}
\usepackage[utf8]{inputenc} 
\usepackage[T1]{fontenc}    
\usepackage{hyperref}       
\usepackage{url}            
\usepackage{booktabs}       
\usepackage{amsfonts}       
\usepackage{nicefrac}       
\usepackage{microtype}      
\usepackage{graphicx}
\usepackage{tabularx}
\usepackage{hyperref}
\usepackage{amsmath}
\usepackage{authblk}

\title{Optimization of Nuclear Mass Models Using Algorithms and Neural Networks}

\author[1]{Jin Li \thanks{Joint first author, Email: 202220102065@mails.zstu.edu.cn}}
\author[1]{Hang Yang \thanks{Joint first author, Email: 202230106326@mails.zstu.edu.cn}}
\affil[1]{\normalsize School of Science, Zhejiang Sci-Tech University, Hangzhou 310018, China}

\begin{document}
\maketitle

\begin{abstract}
Taking into account nucleon-nucleon gravitational interaction, higher-order terms of symmetry energy, pairing interaction, and neural network corrections, a new BW4 mass model has been developed, which more accurately reflects the contributions of various terms to the binding energy. A novel hybrid algorithm and neural network correction method has been implemented to optimize the discrepancy between theoretical and experimental results, significantly improving the model's binding energy predictions (reduced to around 350 keV). At the same time, the theoretical accuracy near magic nuclei has been marginally enhanced, effectively capturing the special interaction effects around magic nuclei and showing good agreement with experimental data.
\end{abstract}

\section{Introduction}\label{sec.I}

The nuclear mass, as one of the fundamental physical properties of atomic nuclei, contains abundant information about nuclear structure \cite{1}. Changes in nuclear mass not only directly affect the stability of atomic nuclei but also play a crucial role in the energy release during nuclear reactions. In particular, for neutron-rich nuclei, their mass is a critical input parameter for the rapid neutron capture process (r-process) in stellar nucleosynthesis processes, and studying it helps in gaining a comprehensive understanding of the formation and evolution of elements in the universe \cite{2,3,4,5}. In recent years, with the ongoing advancements of radioactive ion beam facilities, the masses of over 3000 ground-state nuclei have been experimentally measured, and the experimental measurement scope is gradually extending to both sides of the $\beta$-stability line \cite{6,7}. Since astrophysical research requires a large amount of nuclear mass data for neutron-rich or neutron-deficient nuclei far from the stability line, which remains challenging to measure directly with current technology, many different types of nuclear mass models have been proposed to address the limitations of existing experimental techniques.

The semi-empirical mass formula proposed by Bethe and Weizs?cker in the early days had a mass prediction accuracy of approximately 3 MeV \cite{8,9,10}. The Strutinsky energy theorem \cite{11} states that nuclear binding energy can be divided into two parts: one large and smooth, and the other small and oscillatory. Due to its limitations, the classical liquid drop model can only explain smooth trends but fails to account for the rapid oscillations in binding energy around shell gaps as a function of proton and neutron numbers. This suggests that some important physical effects are missing from the classical model \cite{12,13}. To address this issue, physicists developed macroscopic-microscopic models by introducing shell corrections. These models include the Finite-Range Droplet Model (FRDM) \cite{14}, the Koura-Tachibana-Uno-Yamada (KTUY) model \cite{15}, the Lublin Strasbourg Drop (LSD) model \cite{16}, and the Weizsacker-Skyrme (WS) mass model \cite{17}. In addition, there are microscopic mass models based on Density Functional Theory (DFT) \cite{18}, such as the Hartree-Fock-Bogoliubov (HFB) \cite{19} method and the Relativistic Mean Field (RMF) theory \cite{20}. Although these theoretical models are more complex, they possess better extrapolation capabilities and can describe nuclear mass and structure more accurately.

To address issues such as the lack of physical effects and overfitting in early semi-empirical mass formulas, Professor Kirson's team introduced six physical terms into the model, including the exchange Coulomb term, surface symmetry term, and shell effects term \cite{21,22,23,24,25,26,27}. These physical constraints significantly improved the accuracy of the model, and the resulting BW2 mass model partially addressed the original issues. Furthermore, thanks to the ability of machine learning to handle complex problems, the BW2 model has found wide application in nuclear physics, such as predicting half-lives, charge radii, and charge densities \cite{28,29,30,31,32}.

Neural Networks are an important method in machine learning, inspired by the structure and function of biological neural systems. They consist of a series of interconnected artificial neurons, simulating the brain's learning process by adjusting connection weights. Neural networks are particularly adept at handling complex nonlinear relationships and high-dimensional data, and are thus widely used in fields such as image recognition, natural language processing, and predictive analytics \cite{33}. In the optimization of nuclear mass models, neural networks can effectively capture the complex relationships within nuclear physics data. Traditional nuclear mass models may be limited by parameter selection and model assumptions, while neural networks, through training on large datasets, can adaptively adjust the model structure and parameters, improving accuracy and revealing underlying patterns and trends in nuclear mass \cite{34,35}. By incorporating neural networks to optimize nuclear mass models, not only can the performance of existing models be improved, but new ideas and methods for nuclear physics research can also be provided.

This work is based on the BW2 nuclear mass model. By introducing higher-order terms for symmetry energy, gravitational terms, and pairing interaction terms, combined with neural network-based model corrections, we have developed a new BW4 nuclear mass model. To verify the accuracy and performance of the model and method, we conducted detailed testing and analysis. The structure of the rest of this paper is as follows: In Section 2, we introduce the nuclear mass model, the principles of the algorithm, and the fundamentals of neural networks. In Section 3, we first introduce the evaluation metrics for model performance and analyze the performance improvements of the BW4 mass model under the optimization of the algorithm and neural network, from a local to global perspective. The final section provides a conclusion.

\section{Mass Model and Algorithm}\label{sec.II}
\subsection{BW2 Mass Model}\label{A}

The BW2 mass model \cite{36} is based on the classical liquid drop model, and by adding six physical terms as multiple constraints, it optimizes the deviations of the semi-empirical mass formula to a certain extent. The BW2 mass model is as follows:

\begin{equation}
	\begin{aligned}
		BE_{\text{BW2}} = & \, \alpha_r A + \alpha_s A^{\frac{2}{3}} + \alpha_c \frac{Z^2}{A^{\frac{1}{3}}} + \alpha_t \frac{(N-Z)^2}{A} + \alpha_p \frac{Z^{\frac{4}{3}}}{A^{\frac{1}{3}}} \\
		& + \alpha_{\text{cc}} \frac{|N-Z|}{A} + \alpha_{\text{sx}} \frac{(N-Z)^2}{A^{\frac{4}{3}}} + \alpha_{\text{so}} \delta A^{-\frac{1}{2}} \\
		& + \alpha_{\pi} A^{\frac{1}{3}} + \alpha_m P + \beta_m P^2
	\end{aligned}
\end{equation}

Equation (1) contains 11 fitting coefficients. Here,
\[
P = \frac{v_n v_p}{v_n + v_p},
\]
\[
\delta(N,Z) = \frac{[(-1)^N + (-1)^Z]}{2},
\]
\[
\nu_{p}, \nu_{n}
\]
represent the differences between the actual proton number (Z) and neutron number (N) and the nearest magic numbers, respectively.

\subsection{Improvements in the Mass Model}\label{B}

By considering the Fermi gas model to explain the binding energy of nucleons, we introduced higher-order terms for the symmetry energy
\[
\alpha_{\text{tm}} \frac{(N-Z)^4}{A^3}.
\]
Here,
\[
\alpha_{\text{tm}} = \frac{1}{162} \left(\frac{9\pi}{8}\right)^{2/3} \frac{\hbar^2}{m r_0^2}
\]
represents the modified Planck constant, while \( m \) and \( r_0 \) denote the mass and radius of the nucleon, respectively.

Gravity is the most important interaction between objects on a macroscopic scale, and its range is infinite. Although the influence of gravity weakens as objects move farther apart, the gravitational interaction between nucleons in the liquid drop model cannot simply be ignored. Its expression is defined as:
\[
BE_g = \alpha_g \frac{A(A-1)}{A^{1/3}},
\]
\(\alpha_g\) is the fitting coefficient for the gravitational term.

Additionally, we found that in atomic nuclei, nuclei with an even number of nucleons tend to be more stable than those with an odd number of nucleons. Even-numbered nucleons can pair up, while odd-numbered nucleons cannot fully pair. Therefore, we considered introducing a pairing interaction term to make the overall model more stable. The form of the pairing interaction term is:

\[
BE_{\text{pm}} = \alpha_{\text{pm}} A^{-1/3} \delta_{\text{pm}},\]

\[\quad \delta_{\text{pm}} =
\begin{cases}
	2 - \left| \frac{N - Z}{A} \right|, & \text{N and Z even} \\
	\left| \frac{N - Z}{A} \right|, & \text{N and Z odd} \\
	1 - \left| \frac{N - Z}{A} \right|, & \text{N even, Z odd, N \textgreater Z} \\
	1 - \frac{\left| N - Z \right|}{4A}, & \text{N odd, Z even, N \textless Z} \\
	1, & \text{N even, Z odd, N \textless Z} \\
	1, & \text{N odd, Z even, N \textgreater Z}
\end{cases}
\]

The aforementioned section explains the large and smooth part of the liquid drop model, which is defined as $BE_{\text{LDM}}$.
However, the explanation for the small and oscillatory part is not ideal. Therefore, we introduced a neural network correction term to account for the small and oscillatory part. This term is defined as $\delta_{\text{NN}}$.

After adding the aforementioned terms, we obtained a new mass model, BW4, as shown in Equation (2):
\begin{equation}
	\begin{aligned}
		BE_{\text{BW4}} = & \, \alpha_r A + \alpha_s A^{\frac{2}{3}} + \alpha_c \frac{Z^2}{A^{\frac{1}{3}}} + \alpha_t \frac{(N-Z)^2}{A} + \alpha_p \frac{Z^{4/3}}{A^{1/3}} \\
		& + \alpha_{\text{cc}} \frac{|N-Z|}{A} + \alpha_{\text{sx}} \frac{(N-Z)^2}{A^{4/3}} + \alpha_{\text{so}} \delta A^{-1/2} \\
		& + \alpha_{\pi} A^{1/3} + \alpha_m P + \beta_m P^2 + \alpha_{\text{tm}} \frac{(N-Z)^4}{A^3} \\
		& + \alpha_g \frac{A(A-1)}{A^{1/3}} + \alpha_{\text{pm}} A^{-1/3} \delta_{\text{pm}} + \delta_{\text{NN}}
	\end{aligned}
\end{equation}

\subsection{Algorithm Principles}\label{C}
In this paper, we selected the Broyden-Fletcher-Goldfarb-Shanno (BFGS) algorithm and Sequential Least Squares Programming (SLSQP) for optimizing the coefficients of the BW4 mass model, and compared them with the commonly used least squares method \cite{35,36,37,38,39,40,41,42}.
\subsubsection{SLSQP}\label{a}
For solving constrained optimization problems (COP), SLSQP fully utilizes gradient and Hessian matrix information, allowing it to converge to the optimal solution more quickly. For any COP:

\begin{equation}
	\min_{\vec{x} \in X} f(\vec{x}) \quad
	\text{s.t.} \quad g(\vec{x}) = 0, \quad h(\vec{x}) \geq 0
\end{equation}

\noindent
where $\vec{x} = (x_1, x_2, x_3, \ldots, x_k)$, with $X = \{\vec{x} \mid \vec{l} \leq \vec{x} \leq \vec{u}\}$,
$\vec{l} = (l_1, l_2, l_3, \ldots, l_i)$, and $\vec{u} = (u_1, u_2, u_3, \ldots, u_j)$.
Here, $\vec{x}$ represents the solution vector of the problem, $X$ is the vector space of the solution, $\vec{l}, \vec{u}$ are the lower and upper boundary constraints of the solution space, $g(\vec{x})$ represents the equality constraint, $h(\vec{x})$ represents the inequality constraint, and $f(\vec{x})$ is the objective function to be optimized \cite{43}.
SLSQP finds the minimum of the objective function under constraints through iterative optimization. During each iteration, the gradient and Hessian matrix \cite{42} of the objective function are computed to determine the search direction, and a linear approximation model is employed to update the current solution. Meanwhile, the satisfaction of constraints is also taken into account, and constraints are addressed by introducing Lagrange multipliers:
\[
L(\vec{x}, \vec{\lambda}, \vec{\mu}) = f(\vec{x}) + \vec{\lambda}^T \ast g(\vec{x}) + \vec{\mu}^T \ast h(\vec{x})
\]
Here, the superscript \( T \) denotes the transpose of the vector, \( \vec{\lambda} \) and \( \vec{\mu} \) represent the penalty terms for equality and inequality constraints, respectively \cite{44}.
By solving the unconstrained least squares problem, the update rule for each iteration is obtained. This rule must satisfy not only the equality and inequality constraints but also the first-order necessary conditions:
\[
\nabla L(\vec{x}, \vec{\lambda}, \vec{\mu}) = \nabla f(\vec{x}) + J_g^T \ast \vec{\lambda} + J_h^T \ast \vec{\mu} = 0 \quad \text{(5)}
\]
Here, \( J_g \) and \( J_h \) represent the Jacobian matrices of the equality and inequality constraint functions, respectively \cite{45}.
According to the aforementioned update rule, the initial value \( \vec{x}_1 \) is selected, and the stopping criterion is defined; the gradient vector \( \nabla f_k(\vec{x}_k) \) is calculated, where \( k \) represents the current iteration number. If \( \|\nabla f_k(\vec{x}_k)\| < \epsilon \), the algorithm terminates, yielding the approximate solution \( \vec{x}^* \), where \( \epsilon \) is the predefined stopping criterion. We construct a second-order sequential quadratic programming (SQP) model:
\[
\text{min}[q(\vec{x})] = \text{min} \left[ f_k(\vec{x}) + g_k^T(\vec{x} - \vec{x}_k) + \frac{1}{2}(\vec{x} - \vec{x}_k)^T B_k(\vec{x} - \vec{x}_k) \right]
\]
\[
\text{s.t.} \quad \begin{cases}
	A_{\text{eq}} (\vec{x} - \vec{x}_0) = 0 \\
	g_k(\vec{x}) \geq 0, \, k = 1, 2, \dots, k
\end{cases} \quad \text{(6)}
\]
Here, \( B_k \) is a positive definite symmetric matrix that approximates the inverse of the Hessian matrix, and \( A_{\text{eq}} \) represents the Jacobian matrix of the equality constraints. Solving the sequential quadratic programming (SQP) model provides the correction direction \( \Delta \vec{x} \); subsequently, the step size \( \alpha \) is calculated to guarantee adequate descent of the objective function in the search direction:
\[
\alpha = \text{min}(1, r^c)
\]
\[
r = \text{max}(\beta_s, r_t)
\]
\[
\beta_s = \left( \frac{\partial f}{\partial \vec{x}} \right)^T (\Delta \vec{x} / s)
\]
\[
r_t = \left( \frac{\partial g}{\partial \vec{x}} \right)^T (\Delta \vec{x} / t) \quad \text{(7)}
\]
\( s \) and \( t \) are positive scaling factors.
Finally, the estimated point is updated as \( \vec{x}_{k+1} = \vec{x}_k + \alpha \Delta \vec{x} \). By solving the aforementioned system of equations, the optimal solution for the first iteration can be obtained. Following this iterative process, the objective function is gradually optimized, and the optimal solution that satisfies the constraints is found.

\subsubsection{BFGS}\label{b}
BFGS (Broyden-Fletcher-Goldfarb-Shanno) is a quasi-Newton method for numerical optimization.

For any quasi-Newton equation:
\[
\nabla f(\vec{x}_k) = \nabla f(\vec{x}_{k+1}) + G_{k+1}^* (\vec{x}_k - \vec{x}_{k+1})
\]
By rearranging terms:
\[
G_{k+1}^* (\vec{x}_{k+1} - \vec{x}_k) = \nabla f(\vec{x}_{k+1}) - \nabla f(\vec{x}_k)
\]
Let \( H_{k+1} \approx G_{k+1} \), we get:
\[
H_{k+1}^* (\vec{x}_{k+1} - \vec{x}_k) = \nabla f(\vec{x}_{k+1}) - \nabla f(\vec{x}_k)
\]
It is usually assumed in BFGS that:
\[
H_{k+1} = H_k + E_k
\]
Let's assume that:
\[
E_k = \alpha_k u_k u_k^T + \beta_k v_k v_k^T
\]
Here, \( u_k \) and \( v_k \) are both \( n \times 1 \) vectors:
\[
y_k = \nabla f(\vec{x}_{k+1}) - \nabla f(\vec{x}_k), \quad s_k = \vec{x}_{k+1} - \vec{x}_k
\]
\[
u_k = r H_k s_k, \quad v_k = \theta y_k
\]
Substituting into the original equation, it can be written as:
\[
\Rightarrow \alpha \left( u_k^T s_k \right) u_k + \beta \left( v_k^T s_k \right) v_k = y_k - H_k s_k
\]
\[
\Rightarrow \alpha \left( \left( H_k s_k \right)^T s_k \right) H_k s_k + \beta \left( \left( \theta y_k \right)^T s_k \right) \left( \theta y_k \right) - v_k + H_k s_k = 0
\]
\[
\Rightarrow \alpha \left( \left( H_k s_k \right)^T s_k \right) H_k s_k + \beta \left( \left( \theta y_k \right)^T s_k \right) \left( \theta y_k \right) - \left( y_k - H_k s_k \right) = 0
\]
\[
\Rightarrow \alpha \left( \left( H_k s_k \right)^T s_k \right) \left( H_k s_k \right) + \beta \left( \theta^2 \left( y_k^T s_k \right) + 1 \right) \left( H_k s_k \right) = 0
\]
\[
\Rightarrow \alpha r^2 = - \frac{1}{s_k^T H_k s_k}, \quad \beta \theta^2 = \frac{1}{y_k^T s_k}
\]
\[
H_{k+1} = H_k + \frac{H_k s_k s_k^T H_k}{s_k^T H_k s_k} - \frac{y_k y_k^T}{y_k^T s_k}
\]
The algorithm proceeds through the following steps: first, initialization, then calculating the gradient of the objective function and updating the search direction; after selecting an appropriate step size, the parameters are updated, followed by updating the inverse Hessian matrix. Next, it checks whether the termination condition is satisfied; if not, the iteration continues until the preset conditions are met, and the final result is obtained \cite{46}.

\subsection{Principles of Neural Networks}\label{D}
\subsubsection{KANs}\label{a}
Kolmogorov-Arnold Networks (KANs) are a new type of neural network inspired by the Kolmogorov-Arnold representation theorem, which states that any multivariable continuous function \(f(x_1, x_2, \dots, x_{n-1}, x_n)\) on a bounded domain can be expressed as a finite combination of univariate continuous functions and addition operations:
\[
f(x_1, x_2, \dots, x_{n-1}, x_n) = \sum_{q=1}^{2n+1} \Phi_q \left( \sum_{p=1}^{n} \varphi_{q,p} (x_p) \right)
\]
Here, \(\varphi_{q,p}\) and \(\Phi_q\) represent univariate functions. This implies that multivariable functions can be expressed via univariate functions and addition, reducing the complexity of high-dimensional functions.

KANs consist of multiple layers, where each layer is made up of a set of univariate functions \(\varphi_{q,p}\) parameterized as B-spline curves with trainable coefficients. The connections between layers are not fixed activation functions but are learnable activation functions, represented by B-splines and parameterized as local B-spline basis functions.

\subsubsection{LSTM}\label{b}
Long Short-Term Memory (LSTM) is a specialized type of recurrent neural network (RNN) initially designed for processing and predicting time series data. However, LSTM's unique properties also give it an advantage in non-time series tasks. As an enhanced network, it effectively captures complex dependencies in the input data.

An LSTM network consists of multiple LSTM units, each of which includes a cell state (Cell State), an input gate (Input Gate), a forget gate (Forget Gate), and an output gate (Output Gate). The cell state is the central component of the LSTM unit, responsible for storing information and transmitting it across different time steps (or sequential steps). The gating mechanism governs the update and maintenance of the cell state, enabling the LSTM to effectively remember long-term dependencies.

The LSTM's gating mechanism consists of three gates: the input gate, the forget gate, and the output gate. Each gate is governed by a Sigmoid activation function, whose output values range between 0 and 1, determining the retention or discarding of information \cite{47}.

The forget gate determines how much information should be forgotten from the current cell state. The formula is:
\[
f_t = \sigma \left( W_f \left[ h_{t-1}, x_t \right] + b_f \right)
\]
Here, \(f_t\) is the output of the forget gate, \(\sigma\) is the Sigmoid activation function, \(W_f\) and \(b_f\) represent the weight matrix and bias vector of the forget gate, \(h_{t-1}\) denotes the hidden state from the previous time step, and \(x_t\) is the current input.
The input gate determines which new information should be added to the cell state. The formula is:
\[
i_t = \sigma \left( W_i \left[ h_{t-1}, x_t \right] + b_i \right)
\]
The new candidate information is produced by a \(\tanh\) layer:
\[
\tilde{C}_t = \tanh \left( W_c \left[ h_{t-1}, x_t \right] + b_c \right)
\]
Then, the cell state is updated together with the input gate:
\[
C_t = f_t * C_{t-1} + i_t * \tilde{C}_t
\]
The output gate decides which information will be output from the cell state. The formula is:
\[
O_t = \sigma \left( W_o \left[ h_{t-1}, x_t \right] + b_o \right)
\]
The final output, along with the cell state, is processed through a \(\tanh\) layer:
\[
h_t = o_t * \tanh \left( C_t \right)
\]
When processing non-time series data, LSTM enhances model performance by capturing the sequential dependencies and relationships between the upper and lower parts of the input data.

\section{Results and discussions}\label{sec.III}
We enhanced the overall stability of the model by considering the analysis of binding energy using the Fermi gas model, the gravitational interaction between nucleons, and incorporating a pairing interaction term. Building on the above neural network principles, we trained a neural network model on the residuals of the binding energy to obtain the necessary neural network correction term, ultimately developing the BW4 mass model.

\subsection{Model Performance Metrics}

The performance of the mass model is evaluated based on the root mean square deviation (RMSD), as defined in Equation (\ref{eq:rmsd}):

\begin{equation}
	\label{eq:rmsd}
	\text{RMSD} = \sqrt{\frac{1}{n} \sum_{i=1}^{n} \left( BE_{\text{Ex}_i} - BE_{\text{Th}_i} \right)^2}
\end{equation}

Here, $n$ denotes the total number of nuclides involved in the calculation, and $BE_{\text{Ex}_i}$ and $BE_{\text{Th}_i}$ represent the experimental binding energy and theoretical model value for each nuclide, respectively.

\subsection{\textbf{Optimization of } $BE_{\text{LDM}}$}
For the $BE_{\text{LDM}}$ component of the BW4 mass model, we optimized using multiple algorithms, conducted relevant calculations on the dataset, and fitted the optimal coefficients for each term under the current algorithm, as shown in Figure 1.
\begin{figure}[h]
	\centering
	\includegraphics[width = 8.4cm]{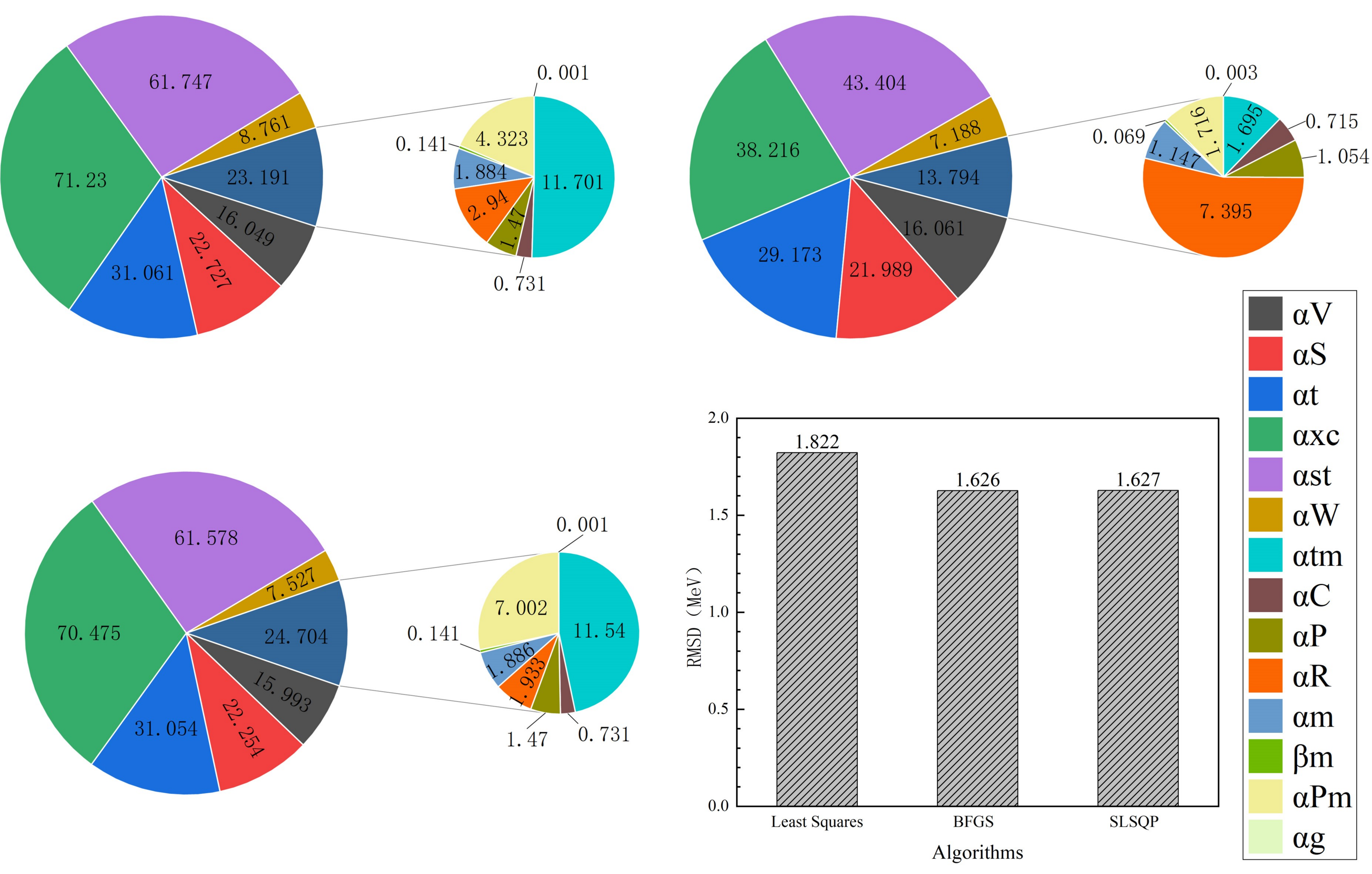}
	\caption{\label{} Optimal coefficients for $BE_{\text{LDM}}$ and RMSD across different algorithms (Unit: MeV)}
\end{figure}

Figure 1 presents the coefficients obtained from algorithms like BFGS and SLSQP. Compared to Least Squares, the other algorithms have caused changes in the weights of the model terms, as reflected in the table. The magnitude of the weights indicates the degree of each term's influence on the overall model, while the sign of each coefficient indicates whether it contributes a positive or negative correction to the model. We focus on the high performance of BFGS (RMSD = 1.626 MeV) and SLSQP (RMSD = 1.627 MeV) in Figure 1. In these two algorithms, the exchange Coulomb term, surface symmetry term, higher-order symmetry energy term, and pairing correction term significantly affect the model, resulting in higher weights, while other terms have smaller impacts and thus lower weights. Hence, these algorithms result in a lower root mean square deviation (RMSD) compared to the others.

Additionally, we focus on the performance of $BE_{\text{LDM}}$ on double magic isotope chains. Figure 2 illustrates the performance of $BE_{\text{LDM}}$ on different isotope chains (Ca, Ni, Sn, and Pb) under various algorithms (BFGS, Least Squares, SLSQP). The x-axis represents the neutron number, and the y-axis shows the relative error between the experimental values and theoretical calculations.

\begin{figure}[h]
	\centering
	\includegraphics[width = 8.4cm]{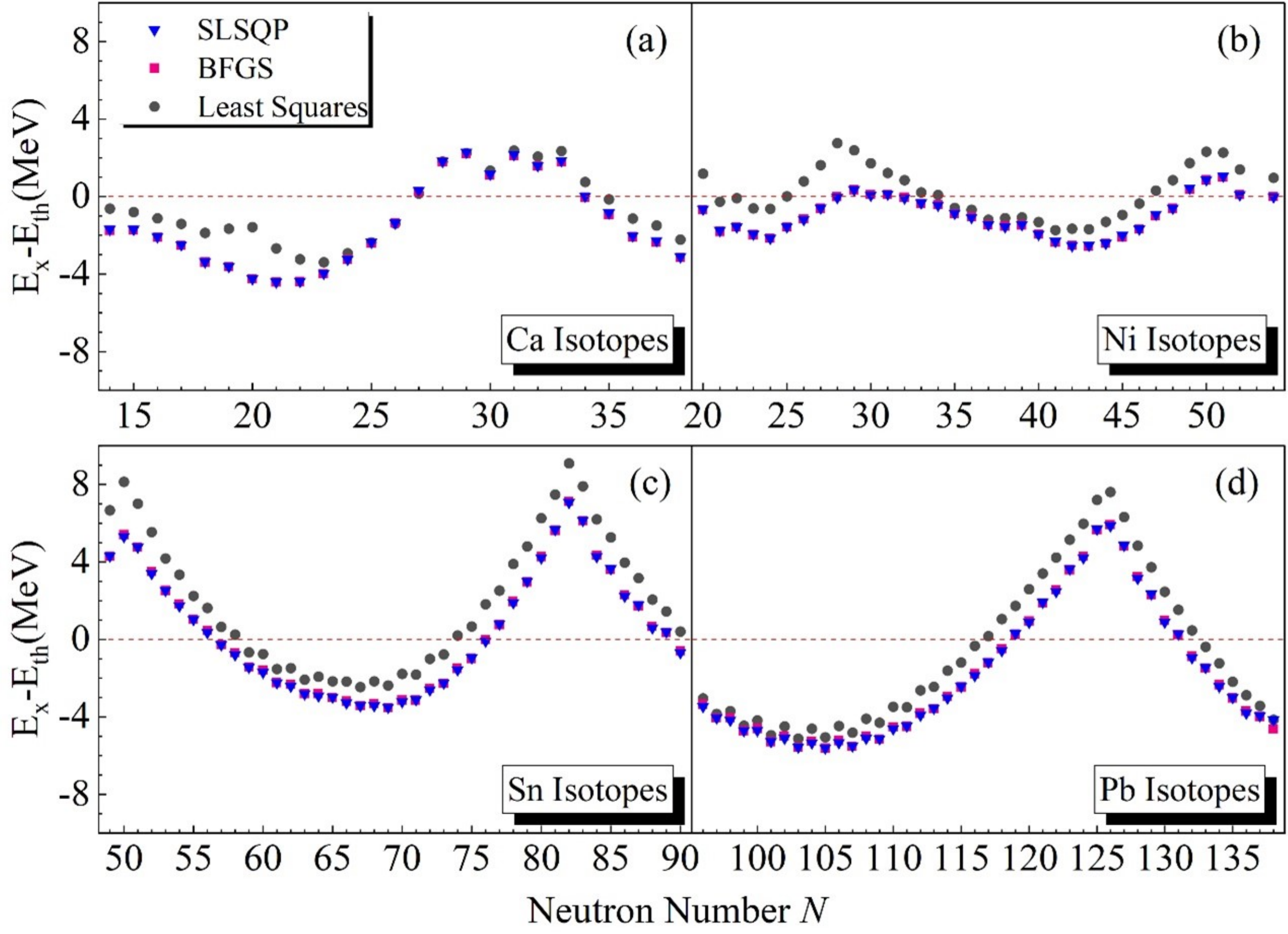}
	\caption{\label{} The relative error between the experimental and theoretical values of $BE_{\text{LDM}}$ for the isotope chains of Ca, Ni, Sn, and Pb.
	}
\end{figure}
Figure 2(a) illustrates the optimization performance of three algorithms along the Ca isotope chain. Overall, the Least Squares algorithm outperforms the SLSQP and BFGS algorithms in terms of deviation from experimental values, while the SLSQP and BFGS algorithms show similar performance. At $N=20$, the error between the Least Squares algorithm and the SLSQP and BFGS algorithms shows a significant divergence from the experimental values. The Least Squares algorithm yields a deviation of -1.579 MeV at $N=20$, whereas the SLSQP and BFGS algorithms produce deviations of -4.257 MeV and -4.243 MeV, respectively. However, at $N=28$, the deviations from the experimental values are minimal across the three algorithms, with the Least Squares, SLSQP, and BFGS algorithms showing deviations of 1.805 MeV, 1.793 MeV, and 1.826 MeV, respectively.

Figure 2(b) presents the optimization of the Ni isotope chain using the three algorithms, where the Least Squares algorithm outperforms the SLSQP and BFGS algorithms before N=28. At N=28, the Least Squares algorithm reaches its largest deviation from the experimental value, at 2.759 MeV, while the deviations for the SLSQP and BFGS algorithms are -0.025 MeV and -0.047 MeV, respectively. At N=50, the SLSQP and BFGS algorithms outperform the Least Squares algorithm, with deviations of 0.878 MeV, 0.865 MeV, and 2.322 MeV, respectively. However, overall, the Least Squares algorithm provides better optimization for the Ni isotope chain, with the SLSQP and BFGS algorithms yielding comparable results.

Figure 2(c) demonstrates the optimization performance of the three algorithms for Sn. For N<57 and N>75, the SLSQP and BFGS algorithms show better performance than the Least Squares algorithm. At N=50 and N=82, the Least Squares algorithm shows its largest deviations from the experimental values, with deviations of 8.137 MeV and 9.091 MeV, respectively. However, for the SLSQP and BFGS algorithms, the deviations from the experimental values at these neutron magic numbers are 5.397 MeV and 7.119 MeV, and 5.32 MeV and 7.076 MeV, respectively. The SLSQP and BFGS algorithms provide better optimization for Sn than the Least Squares algorithm, though their performance remains quite similar to each other.

Figure 2(d) presents the optimization results for Pb using the three algorithms. At N=126, the Least Squares algorithm performs worse than the SLSQP and BFGS algorithms, showing a deviation of 7.606 MeV from the experimental value, while the SLSQP and BFGS algorithms achieve deviations of 5.927 MeV and 5.867 MeV, respectively. However, overall, the Least Squares algorithm demonstrates better optimization performance for Pb compared to the SLSQP and BFGS algorithms, while the performance results between SLSQP and BFGS remain comparable.

\subsection{Optimized $BE_{\text{LDM}}$ Model with Neural Network-Based $\delta_{Nn}$ Correction Term}

In the previous section, the optimized $BE_{\text{LDM}}$ partially reduced the gap between theoretical and experimental values. However, the classical liquid drop model (LDM), due to its inherent limitations, can only capture broad, smooth trends, and is unable to explain the rapid fluctuations in binding energy near shell gaps as a function of proton and neutron numbers. Therefore, we introduce a neural network correction to account for the small-scale fluctuations, enabling the theoretical values predicted by the mass model to better align with experimental data. Accordingly, we incorporate a neural network correction term ($\delta_{Nn}$) to approximate the small, fluctuating components as indicated by the Strutinsky energy theorem.

The dataset was divided into a 7:3 split, consisting of 2275 training samples and 975 testing samples. A neural network model was trained on the binding energy residuals to develop the neural network correction term ($\delta_{Nn}$) for the model. The final performance of the optimized $BE_{\text{LDM}}$ + Neural Network Correction Term ($\delta_{Nn}$) in the BW4 mass model across different datasets is presented in Table 1.
\begin{table}[ht]
	\centering
	\caption{Performance of the BW4 Mass Model Across Datasets (Unit: MeV)}
	\label{tab:performance}
	\begin{tabular}{|c|c|c|c|c|}
		\hline
		\multicolumn{5}{|c|}{\textbf{Full Set (3250 nuclei)}} \\
		\hline
		Unit (MeV) & None & KAN & GPR & LSTM \\
		\hline
		Least Squares & 1.822 & 0.231 & 0.229 & 0.221 \\
		SLSQP         & 1.627 & 0.649 & 0.333 & 0.236 \\
		BFGS          & 1.626 & 0.522 & 0.287 & 0.233 \\
		\hline
	\end{tabular}
	
	\vspace{0.5cm}
	
	\begin{tabular}{|c|c|c|c|c|}
		\hline
		\multicolumn{5}{|c|}{\textbf{Train Set (2275 nuclei)}} \\
		\hline
		Unit (MeV) & None & KAN & GPR & LSTM \\
		\hline
		Least Squares & 1.850 & 0.158 & 0.182 & 0.124 \\
		SLSQP         & 1.643 & 0.625 & 0.280 & 0.150 \\
		BFGS          & 1.642 & 0.483 & 0.245 & 0.155 \\
		\hline
	\end{tabular}
	
	\vspace{0.5cm}
	
	\begin{tabular}{|c|c|c|c|c|}
		\hline
		\multicolumn{5}{|c|}{\textbf{Test Set (975 nuclei)}} \\
		\hline
		Unit (MeV) & None & KAN & GPR & LSTM \\
		\hline
		Least Squares & 1.755 & 0.346 & 0.311 & 0.357 \\
		SLSQP         & 1.591 & 0.701 & 0.431 & 0.364 \\
		BFGS          & 1.589 & 0.604 & 0.366 & 0.352 \\
		\hline
	\end{tabular}
\end{table}

LSTM demonstrated strong performance across the overall dataset when applied with different optimization algorithms. For instance, with the Least Squares optimization algorithm, LSTM achieved a root mean square deviation (RMSD) of 0.221, which was notably lower than other algorithm combinations. SLSQP and BFGS also exhibited strong performance when combined with LSTM, yielding RMSDs of 0.236 and 0.233, respectively. Additionally, KANs and GPR models demonstrated error reduction when paired with different optimization algorithms, though not as pronounced as LSTM. In the training set, the Least Squares + LSTM model exhibited the lowest RMSD of 0.124, demonstrating superior fitting capability. The RMSD values for the SLSQP + LSTM and BFGS + LSTM models were also relatively low, at 0.150 and 0.155, respectively.

The RMSD values for the SLSQP + LSTM and BFGS + LSTM models were also relatively low, at 0.150 and 0.155, respectively. Results from the test set further confirmed these observations. The SLSQP + LSTM and BFGS + LSTM models had errors of 0.364 and 0.352, respectively, also demonstrating strong performance. The KAN and GPR models exhibited relatively higher errors on the test set, highlighting their limitations in generalization ability.

Moreover, the performance of different optimization algorithms varies significantly. The BFGS and SLSQP optimization algorithms, especially when paired with neural network models like LSTM, significantly reduce computational errors in nuclear binding energy, thereby enhancing model accuracy. In contrast, the CG and L-BFGS-B optimization algorithms perform poorly in error control, especially in the absence of model corrections, where they exhibit higher error rates, highlighting their limitations. Neural network models, such as LSTM, effectively enhance the computational precision of nuclear binding energy, bringing theoretical predictions closer to experimental results. Notably, the LSTM model excels in both the training and overall datasets. These findings suggest that the integration of optimization algorithms with neural network models can substantially enhance computational accuracy, offering more precise tools for nuclear physics research.

\begin{figure*}
	\centering
	\includegraphics[width = 14cm]{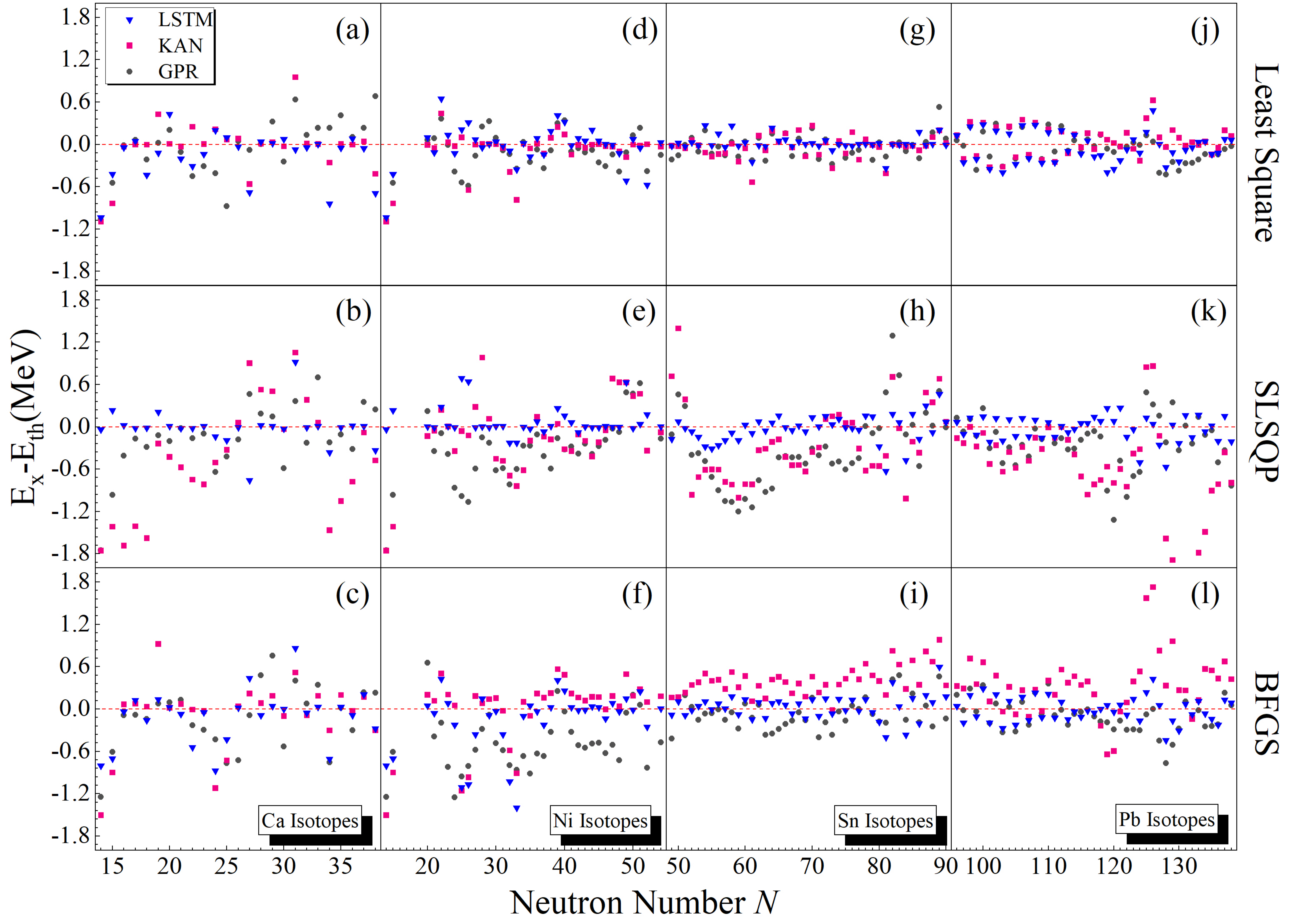}
	\caption{\label{} Relative Errors Between Experimental and Theoretical Values for the BW4 Mass Model Across Ca, Ni, Sn, and Pb Isotope Chains
	}
\end{figure*}

Figure 3 shows the performance of the BW4 model across various isotope chains (Z=20, 28, 50, 82) under different optimization algorithms (BFGS, Least Squares, SLSQP). In these graphs, the horizontal axis represents the neutron number, and the vertical axis indicates the difference between experimental and model-calculated values. These plots reveal the differences in how each algorithm handles the correction of nuclear binding energy.

For the Ca isotope chain, Figure 3\,(a) shows that the error between experimental values and the Least Squares + Neural Network correction term has significantly decreased compared to the version without the neural network correction. Although certain neutron numbers still exhibit noticeable fluctuations, the correction captures key trends in error changes. At $N=20$, the difference between experimental and theoretical values improved from the previous optimal of -1.579 MeV to 0.203 MeV (GPR), 0.009 MeV (KANs), and 0.429 MeV (LSTM). At $N=28$, the difference improved from 1.805 MeV to 0.003 MeV (GPR), 0.011 MeV (KANs), and 0.039 MeV (LSTM).

In contrast, Figure 3\,(b) indicates that the SLSQP + Neural Network correction term exhibits smaller error fluctuations, showing improved stability, though significant error points remain. At $N=20$, the difference between experimental and theoretical values improved from -4.257 MeV to -0.203 MeV (GPR), -0.426 MeV (KANs), and 0.006 MeV (LSTM). At $N=28$, the difference improved from 1.793 MeV to 0.188 MeV (GPR), 0.528 MeV (KANs), and 0.023 MeV (LSTM).

In Figure 3\,(c), the BFGS + Neural Network correction term shows larger error fluctuations across neutron numbers, with pronounced peaks and troughs, indicating less stability for this method on the Ca isotope chain. At $N=20$, the difference between experimental and theoretical values improved from -4.243 MeV to 0.096 MeV (GPR), 0.019 MeV (KANs), and 0.021 MeV (LSTM). At $N=28$, the difference improved from 1.826 MeV to 0.476 MeV (GPR), 0.084 MeV (KANs), and -0.091 MeV (LSTM).

For the Ni isotope chain, Figure 3\,(d) indicates that the error between experimental values and the Least Squares + Neural Network correction term shows a significant reduction compared to the uncorrected version. Although there are still noticeable fluctuations at certain neutron numbers, the correction captures key error trends. At $N=28$, the difference between experimental and theoretical values improved from 2.759 MeV to 0.251 MeV (GPR), 0.007 MeV (KANs), and -0.041 MeV (LSTM). At $N=50$, the difference improved from 2.322 MeV to 0.129 MeV (GPR), -0.005 MeV (KANs), and 0.071 MeV (LSTM).

In contrast, Figure 3\,(e) demonstrates that the SLSQP + Neural Network correction term exhibits smaller error fluctuations, indicating better stability, though significant error points remain. For instance, at $N=28$, the difference between experimental and theoretical values shifted from -0.025 MeV to -0.148 MeV (GPR), 0.981 MeV (KANs), and 0.003 MeV (LSTM). At $N=50$, the difference improved from 0.878 MeV to 0.469 MeV (GPR), 0.435 MeV (KANs), and -0.018 MeV (LSTM).

In Figure 3\,(f), the BFGS + Neural Network correction term shows larger error fluctuations across neutron numbers, with pronounced peaks and troughs, indicating reduced stability for this method on the Ni isotope chain. At $N=28$, the difference between experimental and theoretical values shifted from -0.047 MeV to -0.281 MeV (GPR), 0.081 MeV (KANs), and 0.15 MeV (LSTM). At $N=50$, the difference improved from 0.865 MeV to 0.203 MeV (GPR), 0.188 MeV (KANs), and -0.012 MeV (LSTM).

For the Sn isotope chain, Figure 3\,(i) shows a significant reduction in error between the experimental values and the Least Squares + Neural Network correction term compared to the uncorrected model. Despite some noticeable fluctuations at certain neutron numbers, the correction captures key trends in error changes. At $N=50$, the difference between experimental and theoretical values improved from 8.137 MeV to -0.155 MeV (GPR), -0.028 MeV (KANs), and 0.018 MeV (LSTM). At $N=82$, the difference improved from 9.091 MeV to 0.005 MeV (GPR), 6.31$\times10^{-4}$ MeV (KANs), and -0.004 MeV (LSTM).

In contrast, Figure 3\,(g) demonstrates that the SLSQP + Neural Network correction term exhibits smaller error fluctuations, showing improved stability, though significant error points remain. At $N=50$, the difference between experimental and theoretical values improved from 5.397 MeV to 0.46 MeV (GPR), 1.396 MeV (KANs), and 0.073 MeV (LSTM). At $N=82$, the difference improved from 7.119 MeV to 1.288 MeV (GPR), 0.707 MeV (KANs), and 0.18 MeV (LSTM).

In Figure 3\,(h), the BFGS + Neural Network correction term shows larger error fluctuations across neutron numbers, with pronounced peaks and troughs, indicating reduced stability for this method on the Sn isotope chain. At $N=50$, the difference between experimental and theoretical values improved from 5.32 MeV to 0.145 MeV (GPR), 0.171 MeV (KANs), and 0.104 MeV (LSTM). At $N=82$, the difference improved from 7.076 MeV to 0.418 MeV (GPR), 0.825 MeV (KANs), and 0.38 MeV (LSTM).

For the Pb isotope chain, Figure 3\,(j) demonstrates a significant reduction in the error between experimental values and the Least Squares + Neural Network correction term compared to the uncorrected model, although substantial error fluctuations persist at certain neutron numbers. At $N=126$, the difference between experimental and theoretical values improved from 7.606 MeV to 0.03745 MeV (GPR), 0.62288 MeV (KANs), and 0.48104 MeV (LSTM).

In contrast, Figure 3\,(k) shows that the SLSQP + Neural Network correction term exhibits superior stability, with the smallest error fluctuations and a lower overall error level, demonstrating its strength in handling isotope chains with high proton numbers. At $N=126$, the difference between experimental and theoretical values improved from 5.927 MeV to 0.31665 MeV (GPR), 0.86198 MeV (KANs), and 0.03674 MeV (LSTM).

In Figure 3\,(l), the BFGS + Neural Network correction term's error curve continues to show large fluctuations, suggesting reduced stability for this method on the Pb isotope chain. At $N=126$, the difference between experimental and theoretical values improved from 5.867 MeV to 4.85$\times10^{-4}$ MeV (GPR), 1.72804 MeV (KANs), and 0.42378 MeV (LSTM).

\section{Summary}\label{sec.IV}

The liquid drop model effectively explains the large and smooth components of the binding energy. However, its limitations prevent it from capturing the small, fluctuating components as described by the Strutinsky theorem. We enhanced the BW2 mass model by introducing nucleon gravitational effects, higher-order symmetry energy terms, and pairing interactions ($BE_{\text{LDM}}$), and further optimized BE using algorithms. A neural network correction term ($\delta_{Nn}$) was then added, leading to the development of the BW4 mass model. Based on this model, the following conclusions were reached:

(1) By refining the original BW2 mass model, the root mean square deviation (RMSD) of the $BE_{\text{LDM}}$ was reduced from 1.915 MeV to 1.822 MeV using the Least Squares method. We tested and compared several optimization algorithms, obtaining lower RMSDs of 1.626 MeV (SLSQP) and 1.627 MeV (BFGS). These improvements resulted from the optimized coefficients, which increased the weighting of the exchange Coulomb term, surface symmetry term, higher-order symmetry energy term, and pairing correction term.

(2) In the doubly magic nucleus region, $BE_{\text{LDM}}$ (SLSQP) and $BE_{\text{LDM}}$ (BFGS) outperform $BE_{\text{LDM}}$ (Least Squares); however, in the semi-magic nucleus region, $BE_{\text{LDM}}$ (Least Squares) exhibits smaller errors relative to experimental values.

(3) Subsequently, a neural network correction term was introduced to account for the small and fluctuating components, leading to the development of the BW4 mass model. The BW4 model combined with BFGS+LSTM(KANs) shows a significantly lower overall error compared to the model without the correction term, reducing from 1.822 MeV to 0.233 MeV (0.522 MeV). The BW4 model with BFGS+LSTM provides notable optimization in both the doubly magic nucleus region and the semi-magic nucleus region.

\end{document}